\documentclass[useAMS,usegraphicx,usenatbib]{mn2e}

\newcommand{\Msun}{\ensuremath{\,{\rm M}_\odot}}           
\newcommand{\Teff}{\ensuremath{T_{\rm eff}}}               
\newcommand{\kms}{\,km\,s$^{-1}$}                          
\newcommand{\Apx}{\,\AA\,px$^{-1}$}                        
\newcommand{\apx}{$^{\prime\prime}$\,px$^{-1}$}            
\newcommand{\ion}[2]{{#1}\,{\sc {\small{#2}}}}             
\newcommand{\Porb}{\ensuremath{P_{\rm orb}}}               
\newcommand{\mc}[1]{\multicolumn{2}{c}{#1}}                
\newcommand{\cd}{\ensuremath{\,{\rm cycle\,\,d}^{-1}}}     

\title[SDSS\,J220553.98$+$115553.7 has stopped pulsating]
      {Orbital periods of cataclysmic variables identified by the SDSS.
       IV. SDSS\,J220553.98$+$115553.7 has stopped pulsating}

\author[Southworth, Townsley \& G\"ansicke]
       {John Southworth$^1$\,\thanks{E-mail: j.k.taylor@warwick.ac.uk (JS)},
        D.\ M.\ Townsley$^{2,3}$,
        B.\ T.\ G\"ansicke$^1$,
        \\ $^1$ Department of Physics, University of Warwick, Coventry, CV4 7AL, UK
        \\ $^2$ Department of Astronomy and Astrophysics, University of Chicago, Chicago, IL 60637, USA
        \\ $^3$ Joint Institute for Nuclear Astrophysics, University of Chicago, Chicago, IL 60637, USA
}

\begin{document} \maketitle 

\begin{abstract}
We present time-series VLT spectroscopy and NTT photometry of the cataclysmic variable SDSS\,J220553.98$+$115553.7, which contains a pulsating white dwarf. We determine a spectroscopic orbital period of $\Porb = 82.825 \pm 0.089$\,min from velocity measurements of the H$\alpha$ emission line. A period analysis of the light curves reveals a dominant periodicity at $P_{\rm phot} = 44.779 \pm 0.038$\,min which is not related to the spectroscopic period. However, the light curves do not exhibit a variation at any frequency which is attributable to GW\,Lib-type pulsations, to a detection limit of 5\,mmag. This non-detection is in contrast to previous studies which have found three pulsation frequencies with amplitudes of 9--11\,mmag at optical wavelengths.
Destructive interference and changes to the thermal properties of the driving layer in direct response to accretion can be ruled out as causes of the non-detection. Alternatively, it is feasible that the object has cooled out of the instability strip since a previous (unobserved) dwarf nova superoutburst. This would be the first time this behaviour has been seen in a cataclysmic variable pulsator. Another possibility is that changes in the surface characteristics, possibly induced by accretion phenomena, have modified the surface visibility of the pulsation modes. Further observations, particularly improved constraints on the timescale for changes in the mode spectrum, are needed to distinguish among possible explanations.
\end{abstract}



\section{Introduction}                                                                       \label{sec:intro}

The study of stellar pulsations is a powerful way of constraining the structure and evolution of many types of star \citep{Pamyatnykh99aca}. For white dwarfs (WDs), pulsations can occur at three stages during their evolution  \citep{Winget98jpcm}. The ZZ\,Ceti instability strip is the coolest of these, with effective temperatures (\Teff s) between about 10\,800\,K and 12\,000\,K depending on surface gravity \citep{Mukadam+04apj,Gianninas++05apj}. The pulsation characteristics of ZZ\,Ceti stars depend on where they are within the instability strip \citep{Mukadam+04apj2}. The hotter ZZ\,Cetis generally pulsate with periods of 100-300\,s and amplitudes of up to 3\%, and the cooler ones pulsate with longer periods (up to 1200\,s) and larger amplitudes (up to 30\%).


Cataclysmic variables (CVs) are interacting binaries where a WD is accreting from a less evolved low-mass star \citep{Warner95book,Hellier01book}. A small number of CVs are known to harbour pulsating WDs, the prototype of this class being GW\,Lib \citep{WarnerVanzyl98iaus, Vanzyl+00balta, Vanzyl+04mn}. The pulsation frequencies here hold information on a range of properties which can otherwise be very difficult to obtain for CVs, including the mass and internal structure of the WD, the accretion rate, and the WD rotational velocity and magnetic field strength \citep[e.g.][]{Townsley++04apj}.

\begin{table*} \begin{center}
\caption{\label{tab:obslog} Log of the observations presented in this work.}
\begin{tabular}{lcccccrrr} \hline
Date & Start time & End time & Telescope and & Optical  &  Number of   & Exposure & Mean      \\
(UT) &  (UT)      &  (UT)    &  instrument   & element  & observations & time (s) & magnitude \\
\hline
2007 08 07 & 03:07 & 08:41 & NTT\,/\,SUSI2 & unfiltered  & 438 &     30 & 20.3 \\
2007 08 08 & 04:30 & 09:13 & NTT\,/\,SUSI2 & unfiltered  & 414 & 20--30 & 20.3 \\
2007 08 16 & 05:40 &       & VLT\,/\,FORS2 & $V$ filter  &   1 &     30 & 20.1 \\
2007 08 16 & 05:44 & 07:37 & VLT\,/\,FORS2 & 1200R grism &  15 &    400 &      \\
2007 08 17 & 03:45 &       & VLT\,/\,FORS2 & $V$ filter  &   1 &     20 & 20.2 \\
2007 08 17 & 03:48 & 07:02 & VLT\,/\,FORS2 & 1200R grism &  22 &    480 &      \\
\hline \end{tabular} \end{center} \end{table*}

WDs in accreting binaries are observed to have pulsations for a much wider variety of effective temperatures (\Teff s) than ZZ\,Ceti stars (single WDs), ranging from 10\,500\,K for HS\,2331+3905 \citep{Araujo+05aa} to $\sim$15\,000\,K for several objects \citep{Szkody+07apj}. A theoretical explanation has been put forward by \citet{Arras++06apj}, who found that WDs with a high helium abundance may encounter an additional instability strip at $\Teff \approx 15\,000$\,K due to the deeper convection zone resulting from helium ionization.

\subsection{SDSS\,J220553.98$+$115553.7}


We are currently engaged in a project \citep{Gansicke05aspc} to measure the orbital periods and basic properties of the homogeneous sample of CVs which has been identified by the Sloan Digital Sky Survey (SDSS; \citealt{York+00aj, Szkody+02aj, Szkody+03aj, Szkody+04aj, Szkody+05aj, Szkody+06aj, Szkody+07aj}). Results have so far been presented by \citet{Gansicke+06mn}, \citet{Me+06mn,Me+07mn,Me+07mn2} and \citet{Dillon+08}. As part of this project we have obtained time-resolved photometry and spectroscopy of the pulsating white dwarf system SDSS\,J220553.98$+$115553.7 (hereafter SDSS\,J2205) in order to measure its orbital and pulsation periods.

SDSS\,J2205 is a faint and blue object ($g = 20.05$, $g-r = 0.05$) which was identified as a CV by \citet{Szkody+03aj}, from an SDSS spectrum which shows broad double-peaked Balmer emission lines. The presence of wide absorption features surrounding these emission lines indicates that the WD component contributes a significant proportion of the total light of the system. This is a prerequisite for observing brightness variations arising from WD pulsations.


\citet{WarnerWoudt04asp} presented high-speed unfiltered optical photometry of SDSS\,J2205 which show periodicities of 575\,s, 475\,s and 330\,s, with amplitudes of 9--11\,mmag. \citet{Szkody+07apj} confirmed the presence of the lowest-frequency pulsation, finding a period of $576.2 \pm 1.6$\,s and an amplitude of $46 \pm 11$\,mmag from ultraviolet observations obtained with the {\it Hubble Space Telescope}. From a three-component fit to the spectral energy distribution of SDSS\,J2205, \citet{Szkody+07apj} found a WD \Teff\ of $15\,000 \pm 1000$\,K and a distance of 810\,pc. This \Teff\ is consistent with temperatures found for other GW\,Lib stars \citep{Szkody+02apjl} and with the theoretical predictions of \citet{Arras++06apj}.


\section{Observations and data reduction}                                           \label{sec:obs}

Throughout our observations the apparent magnitude of SDSS\,J3305 was always found to be close to 20.1, which is in agreement with the SDSS imaging and spectroscopic data. This object has therefore never been observed to have large-scale brightness variations.

\subsection{VLT spectroscopy}                                                       \label{sec:obs:vltspec}
                                                                                    \label{sec:obs:vltphot}

Spectroscopic observations were obtained in 2007 August using the FORS2 spectrograph \citep{Appenzeller+98msngr} at the Very Large Telescope (VLT), Chile (Table\,\ref{tab:obslog}). The 1200R grism was used, giving a wavelength interval of 5870\,\AA\ to 7370\,\AA, with a reciprocal dispersion of 0.73\Apx\ and a resolution of 1.6\,\AA\ at H$\alpha$.

The data were reduced using optimal extraction \citep{Horne86pasp} as implemented in the {\sc pamela}\footnote{{\sc pamela} and {\sc molly} were written by TRM and can be found at \\ {\tt http://www.warwick.ac.uk/go/trmarsh}} code (\citealt{Marsh89pasp}), which also makes use of the {\sc starlink}\footnote{Starlink software can be accessed from \\ {\tt http://starlink.jach.hawaii.edu/}} packages {\sc figaro} and {\sc kappa}. Wavelength calibration was performed using one arc lamp exposure for each night, and shifts due to spectrograph flexure were measured and removed using the 6300.304\,\AA\ night sky emission line (see \citealt{Me+06mn}).

The spectroscopic observing procedure of FORS2 included obtaining target acquisition images from which photometry can be obtained. We have extracted differential photometry from these images using the {\sc starlink} package {\sc gaia}. The $V$-band apparent magnitude of the comparison star was calculated from its $g$ and $r$ magnitudes using the transformations provided by \citet{Jester+05aj}.

\subsection{NTT photometry}                                                         \label{sec:obs:ntt}

Light curves were obtained in 2007 August (Table\,\ref{tab:obslog}) using the New Technology Telescope (NTT) at ESO La Silla, Chile, and the SUSI2 CCD mosaic imager \citep{Dodorico+98spie}. CCD\#45 was used, binned by factors of three in both directions to give a resolution of 0.24\apx. The observations were performed without a filter in order to maximise the throughput; this may reduce the pulsation amplitudes (which are larger at bluer wavelengths) but will not smear out the pulsations as they are coherent over all optical wavelengths \citep{Robinson++82apj}. Care was taken to place the target and comparison stars on parts of the CCD which were least affected by fringing. Exposure times of 20--30\,s gave an observing cadence of 36--46\,s.

Debiasing and flat-fielding of the raw images was performed with the {\sc starlink} software packages {\sc convert} and {\sc kappa}. Optimal differential photometry \citep{Naylor98mn} was obtained from the reduced images with the {\sc multiphotom} script \citep{Me++04mn}, which uses the {\sc starlink autophotom} package \citep{Eaton++99}. We adjusted this to an apparent magnitude scale using the mean of the $g$ and $r$ magnitudes of the comparison star. Photometry was also obtained for a check star and used to confirm that the comparison star is not variable.

The resulting light curves show a clear variation with airmass, arising from the use of unfiltered photometry and comparison stars which are significantly redder than SDSS\,J2205. A straight line was fitted to the trend of magnitude versus airmass, and was removed from the light curves.


\section{Results}                                                                    \label{sec:results}

\begin{figure} \includegraphics[width=0.48\textwidth,angle=0]{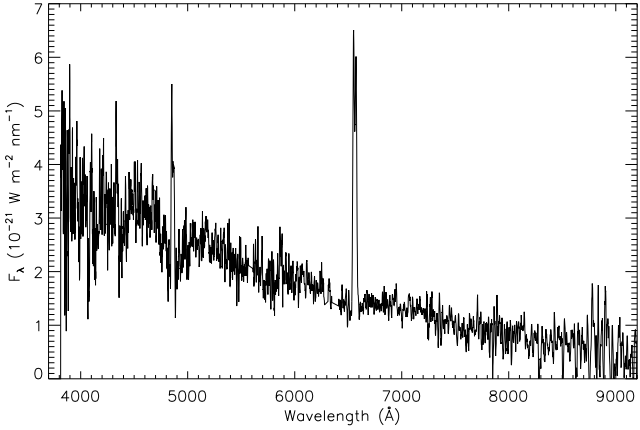} \\
\caption{\label{fig:sdssspec} SDSS spectrum of SDSS\,J2205. For this plot
the flux levels have been smoothed with a 10-pixel Savitsky-Golay filter.
The units of the abscissa are $10^{-21}$\,W\,m$^{-2}$\,nm$^{-1}$, which
corresponds to $10^{-17}$\,erg\,s$^{-1}$\,cm$^{-2}$\,\AA$^{-1}$.} \end{figure}

\begin{figure} \includegraphics[width=0.48\textwidth,angle=0]{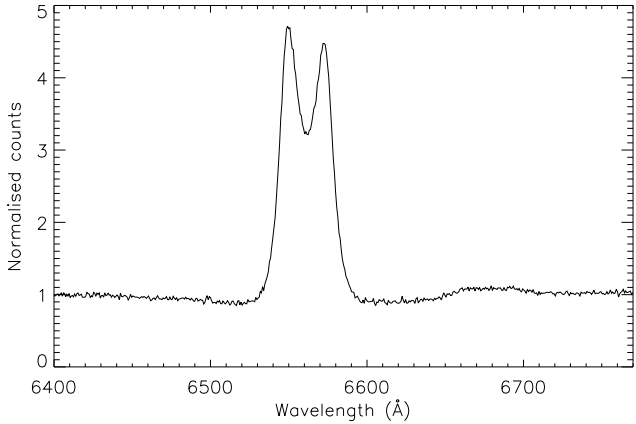} \\
\caption{\label{fig:Halpha} The averaged H$\alpha$ emission line profile
from our continuum-normalised VLT data. The \ion{He}{I} 6678\,\AA\ emission
line is detectable.} \end{figure}

\begin{figure} \includegraphics[width=0.48\textwidth,angle=0]{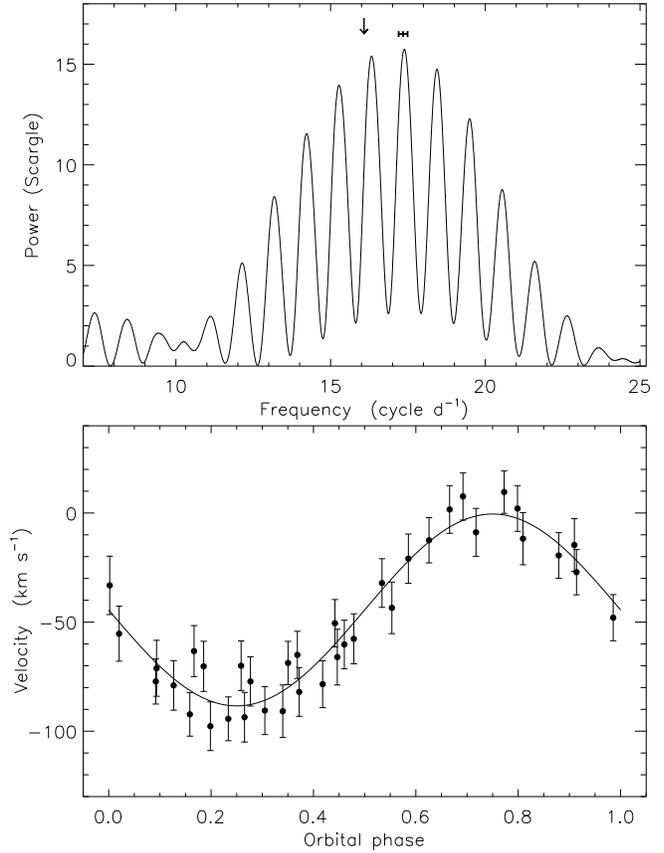}\\
\caption{\label{fig:2205:rvplot} {\it Upper panel:} Scargle periodogram of
the radial velocities of SDSS\,J2205 measured using a double Gaussian with
FWHMs 300\kms\ and separation 1900\kms. The period and $\pm$1$\sigma$
uncertainties from the preliminary period are shown as a horizontal
errorbar, and an arrow indicates half the photometric period.
{\it Lower panel:} measured radial velocities (filled circles) compared
to the best-fitting spectroscopic orbit (unbroken line).} \end{figure}

\subsection{Spectroscopic analysis}                                            \label{sec:results:spec}


The SDSS spectrum of SDSS\,J2205 (Fig.\,\ref{fig:sdssspec}) shows a faint blue continuum with double-peaked Balmer emission lines. Whilst the Balmer absorption of the WD is noticeable, there are no identifiable features arising from the low-mass secondary star. Our VLT spectra (Fig.\,\ref{fig:Halpha}) shows that \ion{He}{I} 6678\,\AA\ emission is weak but detectable. The H$\alpha$ emission equivalent width is $135 \pm 15$\,\AA\ in the SDSS spectrum and $140 \pm 3$\,\AA\ in the mean VLT spectrum.

We have measured radial velocities from this line by cross-correlation against single and double Gaussian functions \citep{SchneiderYoung80apj,Shafter83apj}, as implemented in {\sc molly}. The radial velocities were searched for periods using periodograms computed by the \citet{Scargle82apj} method, as implemented within the {\sc tsa}\footnote{\scriptsize\tt http://www.eso.org/projects/esomidas/doc/user/98NOV/volb/node220.html} context in {\sc midas}. Spectroscopic orbits were fitted using the {\sc sbop}\footnote{Spectroscopic Binary Orbit Program, written by P.\ B.\ Etzel, \\ {\tt http://mintaka.sdsu.edu/faculty/etzel/}} program, which we have found to give reliable error estimates \citep{Me+05mn}.

We first measured radial velocities from the 22 spectra obtained on the night of 2007 August 17, as these were obtained consecutively during a 196-min interval so yield a periodogram which is not affected by cycle count ambiguities. Consistent results were found for all measurement methods, giving a preliminary period of $83.04 \pm 0.70$\,min (using a double Gaussian with widths 300\kms\ and separation 1900\kms).

When we add in the 15 spectra taken on the previous night (2007 August 16) we obtain periodograms with the usual pattern of peaks at the correct orbital period and its one-day aliases. The three highest peaks correspond to periods of 78.0, 82.7 and 88.1 min. The 82.7\,min peak represents the orbital period, as the adjacent possibilities both differ by over 7$\sigma$ from the preliminary period. We therefore find a final orbital period of $\Porb = 82.825 \pm 0.089$\,min, which is close to the minimum period for CVs with relatively unevolved secondary stars \citep{Knigge06mn}.

\begin{figure}
\includegraphics[width=0.235\textwidth,angle=0]{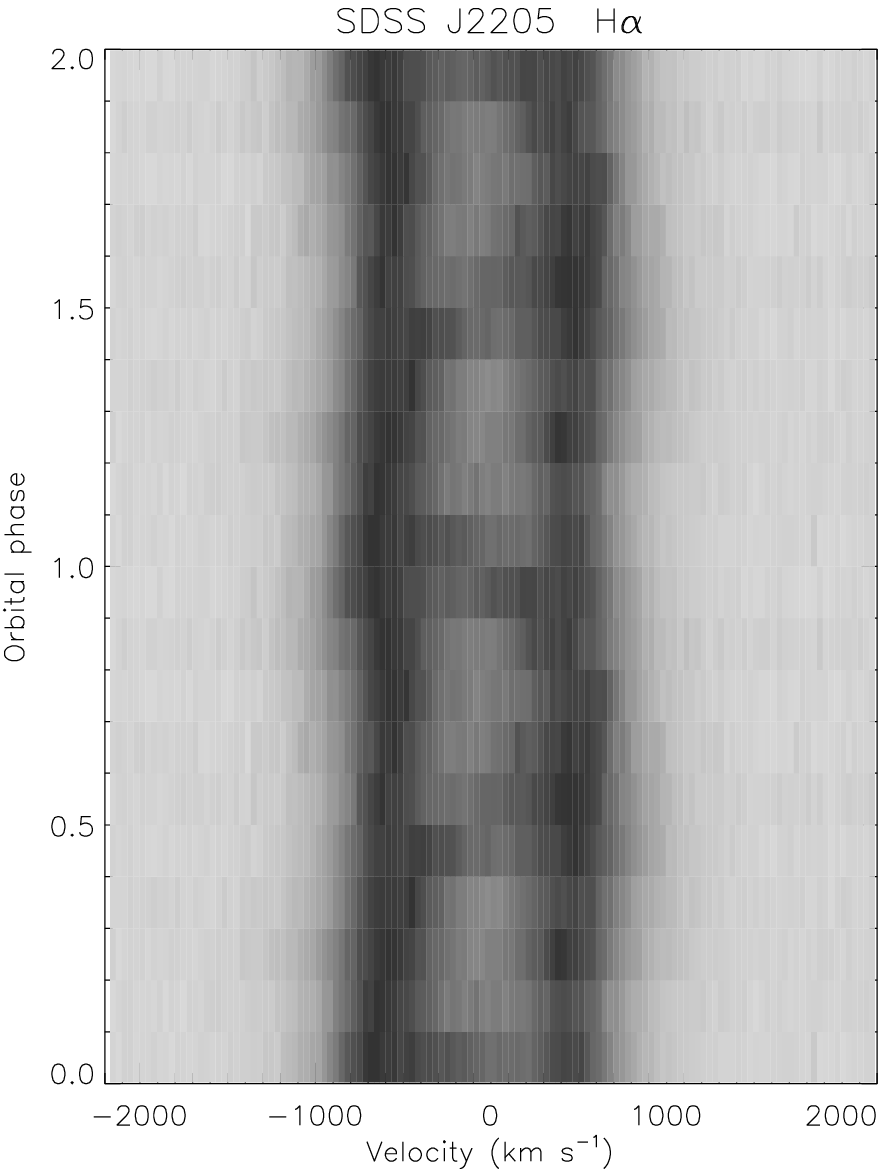}
\includegraphics[width=0.235\textwidth,angle=0]{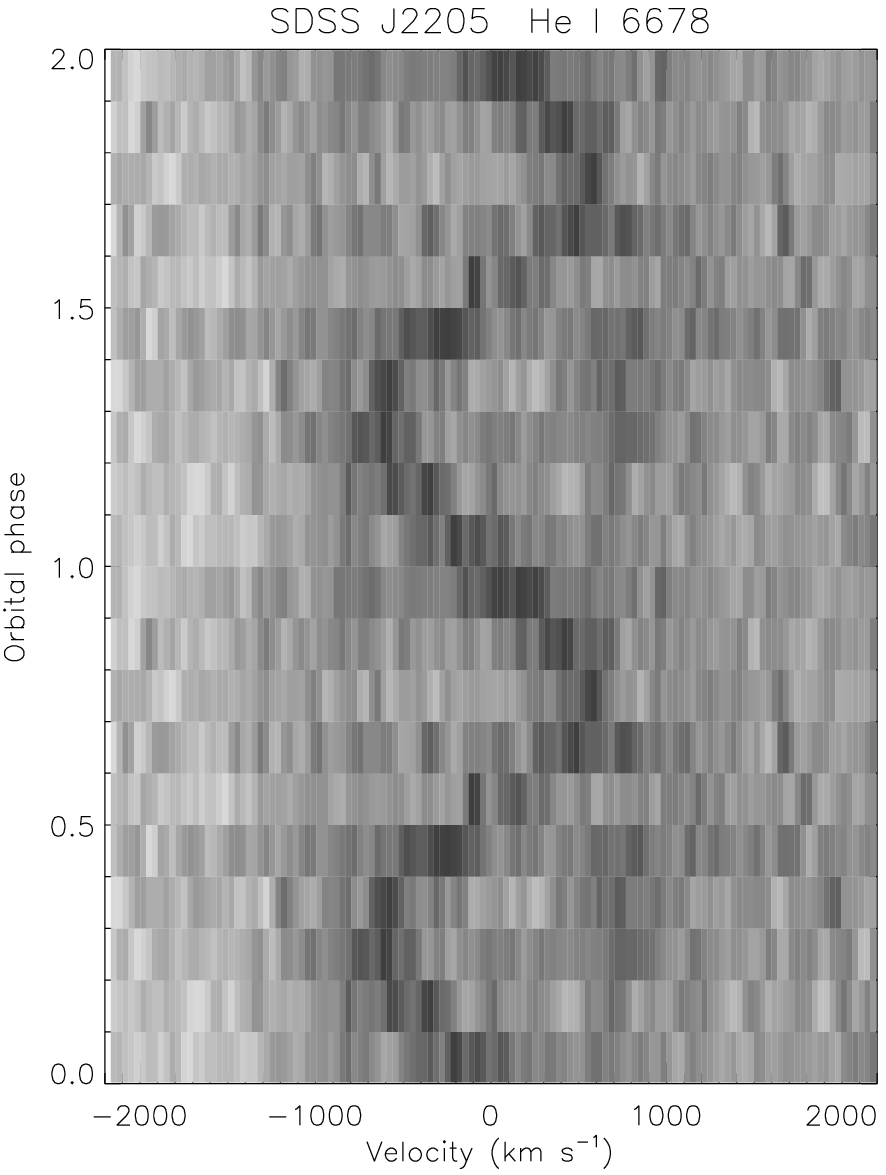}
\caption{\label{fig:trailed} Greyscale plots of the continuum-normalised
and phase-binned trailed spectra around the H$\alpha$ and He\,I 6678\,\AA\
emission lines. The binned spectra for He\,I have been smoothed with a
Savitsky-Golay filter for display purposes.} \end{figure}

The parameters of the spectroscopic orbit are given in Table\,\ref{tab:orbit}. The periodogram and spectroscopic orbit from our full radial velocity dataset are plotted in Fig.\,\ref{fig:2205:rvplot}. We have sorted the spectra into ten phase bins, and Fig.\,\ref{fig:trailed} shows the trailed spectra around the H$\alpha$ and \ion{He}{I} 6678\,\AA\ emission lines. The H$\alpha$ profile is clearly double-peaked, with a fainter S-wave arising from the bright spot on the edge of the accretion disc \citep{Smak85aca}. The \ion{He}{I} 6678\,\AA\ trailed spectrum has been smoothed to make the features of this faint line clearer, and shows very faint double peaks with a brighter S-wave superimposed.

\begin{table} \begin{center}
\caption{\label{tab:orbit} Circular spectroscopic orbit found
for SDSS\,J2205 using {\sc sbop}. The reference time (defined
to indicate zero phase) corresponds to inferior conjunction
of the secondary star.}
\begin{tabular}{l r@{\,$\pm$\,}l} \hline
Quantity                      &         \mc{Value}        \\
\hline
Orbital period (d)            &     0.0575175 & 0.000062  \\
Reference time (HJD)          & 2454329.71251 & 0.00063   \\
Velocity amplitude (\kms)     &          44.0 & 2.3       \\
Systemic velocity (\kms)      &       $-$44.4 & 1.7       \\
$\sigma_{\rm rms}$ (\kms)     &          \mc{9.8}         \\
\hline \end{tabular} \end{center} \end{table}

\subsection{Photometric analysis}

\begin{figure} \includegraphics[width=0.48\textwidth,angle=0]{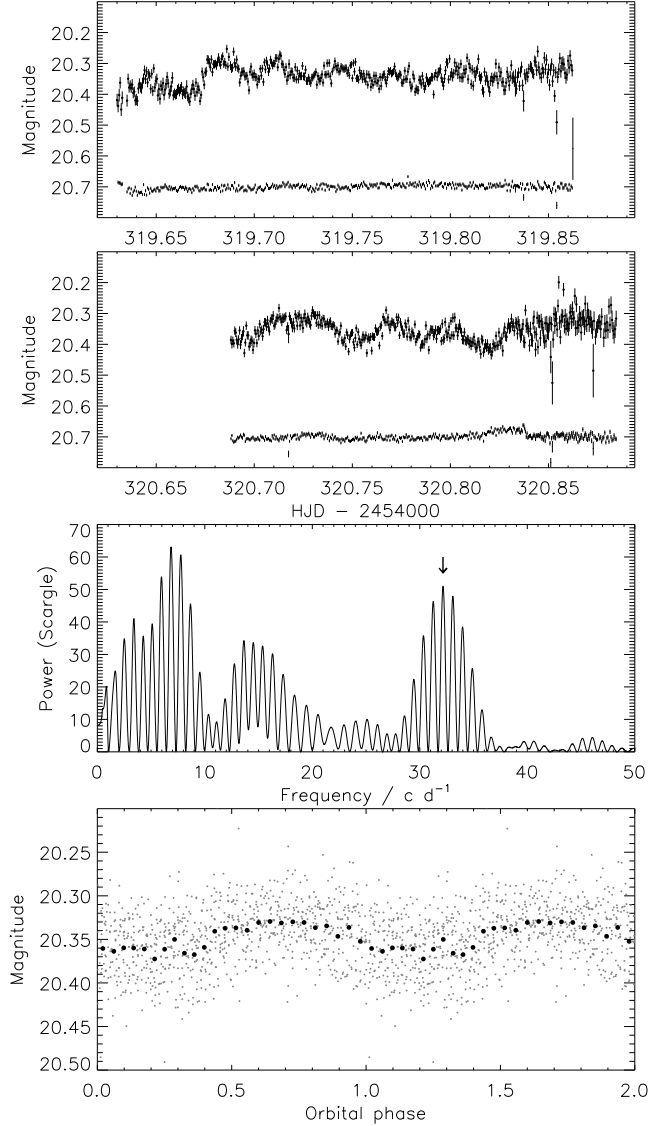}\\
\caption{\label{fig:2205:lcplot} {\it Top panels:} NTT unfiltered light curve
obtained over two nights. The higher light curve in each panel is SDSS\,J2205
minus comparison, and the lower curve is comparison minus check (offset). The
light curves of SDSS\,J2205 have had their variation with airmass removed. The
airmass term for the comparison star light curves is negligible, as it is a
similar colour to the check star, and has not been removed. {\it Panel 3:}
Scargle periodogram of the light curve with the most probable period indicated
with an arrow. {\it Panel 4:} light curve phased at the most probable period
(grey dots) and resampled into 25 phase bins (black filled circles).} \end{figure}

\begin{figure} \includegraphics[width=0.48\textwidth,angle=0]{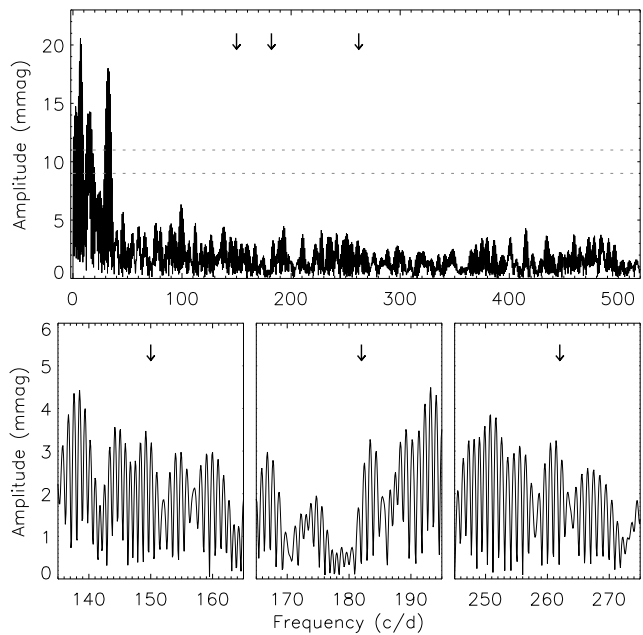}\\
\caption{\label{fig:2205:pplot} {\it Upper panel:} Amplitude spectrum of the NTT
light curve of SDSS\,J2205. The periodicities found by \citet{WarnerWoudt04asp}
and \citet{Szkody+07apj} are indicated with arrows. The dashed lines denote the
amplitudes of these pulsations in the optical observations of
\citet{WarnerWoudt04asp}. {\it Lower panels:} Close-up plots in the regions of
the three periodicities found by other authors.} \end{figure}

We obtained ten hours of high-speed unfiltered photometry over two nights (Section\,\ref{sec:obs:ntt}). The light curves were searched for periods using \citet{Scargle82apj}, AoV \citep{Schwarzenberg89mn} and ORT \citep{Schwarzenberg96apj} periodograms. Fig.\,\ref{fig:2205:lcplot} shows the light curves of SDSS\,2205 and light curves of the comparison star relative to the check star.

We found no significant periodicities in the NTT light curve in the range expected for GW\,Lib-type pulsations. A number of small peaks were found in the AoV and ORT periodograms at higher frequencies (500--1000\cd), but plots of the phased light curve showed that these were all related to the observing cadence rather than to real signals. The amplitude spectrum of the NTT light curve is shown in Fig.\,\ref{fig:2205:pplot} over the range 0--500\cd. The periodicities at 150, 182 and 262 \cd\ found by \citet{WarnerWoudt04asp} and \citet{Szkody+07apj} are indicated with arrows. Magnified plots around these frequencies show no significant power and confirm that the pulsations detected in previous works are not clearly present in our data.

We have determined our detection limit by adding sine waves (frequencies 100--300\cd) to the NTT light curve and recalculating periodograms. We find that sine waves with amplitudes of 5\,mmag or greater would be reliably detected. This detection limit applies specifically to a pulsation which is coherent and persists through the full dataset with a sinusoidal shape. If any of these conditions are not satisfied then the detection limit would increase. The pulsations found by \citet{WarnerWoudt04asp} had amplitudes of approximately 10\,mmag, which would easily have been detected by our observations.

The NTT light curve does show a clear sinusoidal photometric variation with a period of about 45\,min, which is much longer than expected for ZZ\,Ceti-like pulsations \citep{Mukadam+04apj2}. We have calculated periodograms for the individual nights and for the data from both nights. The individual nights give unambiguous periods of $44.71 \pm 0.40$ and $44.09 \pm 0.49$ min. The periodogram for the combined dataset, however, has a large number of peaks, including many at low frequencies which are related to the duration of observations on individual nights  (Fig.\,\ref{fig:2205:lcplot}). The two highest alias peaks correspond to periods of 44.8\,min and 43.5\,min; the first of these is within 1\,$\sigma$ of the mean of the two period measurements from the two individual nights, and the second differs by 3\,$\sigma$. We therefore identify the first periodicity as dominant variation in our light curves. A sine fit gives a final photometric period of $P_{\rm phot} = 44.779 \pm 0.038$\,min.


\section{How could pulsation amplitudes vary?}                                          \label{sec:discussion}

Non-radial pulsations have previously been observed in the WD in SDSS\,J2205 on two occasions \citep{WarnerWoudt04asp,Szkody+07apj}, but were not detectable in our own observations. It is possible that amplitude variation could have taken the pulsations beyond our detection limit, which requires a physical explanation.

Given the current sparse observational coverage, it is plausible that the mode amplitude variations observed are induced by changes in driving. The timescales of amplitude variation in SDSS\,J2205 and other similar objects which have several epochs of observations (e.g.\ SDSS\,J0131, \citealt{Szkody+07apj}; GW\,Lib, \citealt{Vanzyl+04mn}) appear to be months to years. Driving and damping rates in the instability strip for hydrogen-envelope (DA) WDs have been calculated to be around $10^{-6}$\,s$^{-1}$ to $10^{-7}$\,s$^{-1}$ \citep{WuGoldreich99apj}. There are some differences due to the solar chemical composition in the envelopes of DA WDs in CVs, but these rates should have a similar order of magnitude. Thus it is possible that subtle changes in the driving and damping characteristics could explain order unity variations of the driven modes on the timescales relevant for SDSS\,J2205. Such variations in the properties of the driving zone have several possible physical origins, which we now discuss.

\subsection{Changes in accretion?}

The most obvious source of possible variations in driving the pulsations, changes in accretion, is also the most challenging from a physical viewpoint.
By virtue of the nature of the driving mechanism, the thermal timescale of the driving region, on which its temperature structure can change, is similar to the mode period \citep{GoldreichWu99apj} and therefore short compared to the driving timescale.  As a result, in the absence of significant heat sources in this region, the thermal profile in the driving region is set by the heat flux rising from the underlying layers having longer thermal timescales.
The surface luminosity of the WD in accretion quiescence derives from the compression of the accreted material throughout the accreted envelope \citep{TownsleyBildsten04apj}. Only a very small outer portion of the envelope can respond on timescales of months to years, and this portion contributes only a small fraction of the energy flux streaming out through the mode-driving region \citep{TownsleyGansicke08}. Thus only large accretion events, such as superoutbursts, should be able to modify the thermal profile in the driving region. In such a case enough material is accreted that the driving region and layers with longer thermal timescales are replaced by material accreted at a high rate. This will then take some time to cool back to the state defined by the flux from the deepest layers \citep{Piro++05apj}. Such an event would have either been noticed directly or, failing that, would show up as a significant change in the overall system brightness with respect to previous epochs, which we do not observe.

While variation in direct response to accretion is therefore unlikely, this leaves the possible explanation that SDSS\,J2205 had, at the time of our observations, cooled enough since its last superoutburst to leave the instability strip, but only future epochs of observations will tell. If true, this would be an exciting first for CV pulsators. The short-period CV WZ\,Sge had a superoutburst in 2001 July, and observations up to three years after this cataclysmic event have shown that the WD is still cooling \citep{Long+04apj,Godon+06apj}.

Another way to heat the WD surface is via infall heating arising directly from the impact of fresh material. The accretion rate necessary to raise the \Teff\ above that which is set by the underlying flux from envelope compression is only a few percent of the time-averaged accretion rate \citep{TownsleyBildsten03apj}. Thus it is possible to temporarily raise the WD \Teff\ with an episode of fairly mild accretion, which, even if it were months long, might go unnoticed in a dim object like SDSS\,J2205. However, this type of heating \emph{does not affect the driving}, because it is unable to penetrate to the driving region. The temperature in the driving region is $\sim 10^5$\,K so mild increases in the photospheric temperature (which would na{\"\i}vely appear to push the WD out of the instability strip) in fact have no significant effect on the thermal profile at the depths where energy is being pumped into the pulsation modes.

\subsection{Changes in chemical composition?}

The impure composition in the atmospheres of WDs in CV is an essential ingredient for the mode-driving features observed in these systems \citep{Arras++06apj}, particularly the enhanced helium abundance. This means that changes in the composition of the driving layers will be reflected in the balance between driving and damping of modes. If accretion were to cease completely, the timescale for helium to settle out of the convection zone and thus the driving layer for $\Teff \approx 14\,000$\,K is of the order of months \citep{Paquette+86apjs}. However the rate of accretion of material with a solar-like composition need only be $\sim 10^{-14}$\Msun\,yr$^{-1}$ to maintain a similar composition at the driving layer. Thus changes in the quiescent accretion have the potential to induce compositional variation in the driving layers on the timescales of interest. However, it is not clear that such variations would be enough to cause significant changes in mode driving.
More detailed study would be necessary to characterize the time dependence of the chemical structure of the driving layers, and the concomitant impact on the driving.  Additionally, without quantitative constraints on the properties of quiescent accretion, which are difficult due to the low accretion rates involved, conclusions will remain ambiguous even with detailed modelling.

\subsection{Changes in detectability?}

Variations in mode amplitude and therefore driving are not the only viable explanation for the observed amplitude variations. Another possibility is variations in the visibility of any given pulsation mode. Under the influence of rotation each mode will have a unique surface eigenfunction \citep{Bildsten++96apj}. Additionally, the flux perturbations of different modes are modified by the non-adiabatic surface layers in a way which depends on period and the presence of other modes \citep{Brickhill92mn}. In light of the changes in these surface layer which can be achieved by variations in quiescent accretion and the possibility of obscuration by that accretion or other system components, a number of system features could influence the visibility of any given mode. This type of effect has a much wider variety of timescales over which observed amplitude variations could occur. More continuous monitoring of pulsation properties would provide constraints on whether this type of mechanism is at work in these systems.


\section{Summary}                                                   \label{sec:summary} \label{sec:conclusion}

In the context of our program to measure the basic properties of the CVs identified by the SDSS, we present VLT spectroscopy and NTT photometry of the faint short-period system SDSS\,J220553.98+115553.7. This object has twice previously been observed to exhibit low-amplitude brightness variations characteristic of nonradial pulsations \citep{Szkody+07apj,WarnerWoudt04asp}. Using radial velocities measured from the H$\alpha$ emission line in the VLT spectra we find a high-precision orbital period of $\Porb = 82.825 \pm 0.089$\,min, which is close to the observed minimum period for CVs with hydrogen-rich secondary stars.



Our light curves of SDSS\,J2205 cover ten hours over two consecutive nights with an observing cadence of 45\,s. They have a clear sinusoidal variation on a period of $P_{\rm phot} = 44.779 \pm 0.038$\,min. This period is not clearly related to the orbital period or to pulsations, and has not been detected in previous observations of SDSS\,J2205. Unexplained periodicities have been observed in other CVs with pulsating WDs, for example GW\,Lib \citep{WoudtWarner02apss}, HS\,2331+3905 \citep{Araujo+05aa} and SDSS\,J133941.12+484727.5 \citep{Gansicke+06mn}. SDSS\,J2205 is unusual in this company, in that its photometric period is substantially shorter than its orbital period.



We have not found any periodicities in the light curves which are clearly attributable to GW\,Lib pulsations, to a detection limit of 5\,mmag in amplitude. This is in conflict with the optical pulsation amplitudes of 9--11\,mmag found by \citet{WarnerWoudt04asp}. We also have no evidence for significant changes in brightness over longer timescales, which might otherwise have indicated that the WD was contributing a smaller proportion of the light of the system. At the two epochs of the SDSS survey (imaging and spectroscopic), over the four nights of our own observations, and during one night of data presented by \citet{WarnerWoudt04asp}, SDSS\,J2205 has always been found near magnitude 20.1.

There are several potential explanations for the lack of observable pulsations in SDSS\,J2205. Firstly, it is possible for destructive interference to make pulsations unobservable for a few hours \citep{Castanheira+06aa}, but this would be extremely unlikely to occur for ten hours spread over two nights.
A second possibility is cessation or changes in driving due to modification of the thermal state of the driving region. Such changes have several possible origins. If pulsations do not return in future observations, it is likely that, in cooling since its last dwarf nova superoutburst, the object has just left the instability strip.  This behaviour is expected for CV pulsators, but this would be the first time it has been observed. However, it is entirely possible that the current absence of pulsations is a temporary state. Modification of thermal properties of the driving region due to variation in the accretion rate can be excluded because a large increase in accretion would be required, whereas no change in the brightness of the system was observed. Accretion heating of the surface of the WD is also not viable, since it does not affect the driving layer. Changes in driving might also arise from changes in the chemical composition of the driving layer. The plausability of such changes depends on an extremely low quiescent accrection rate, which is difficult to justify. After changes in driving, a third distinct possibility for variation is changes in the visibility of pulsation modes over the surface of the star.  Distinguishing between the several possible origins for pulsation amplitude variation depends on new observations of systems exhibiting pulsations. Most essential for this purpose are constraints on the timescale over which the variations take place, which necessitates more frequent epochs of observation than have been taken to date.

\section*{Acknowledgements}

The reduced spectra, light curves and  radial velocity measurements presented in this work will be made available at the CDS ({\tt http://cdsweb.u-strasbg.fr/}) and at \\ {\tt http://www.astro.keele.ac.uk/$\sim$jkt/}

Based on observations made with ESO Telescopes at the La Silla and Paranal Observatories under programme ID 079.D-0024.

JS acknowledges financial support from PPARC in the form of a postdoctoral research assistant position. The following internet-based resources were used: the ESO Digitized Sky Survey; the NASA Astrophysics Data System; the SIMBAD database operated at CDS, Strasbourg, France; and the ar$\chi$iv scientific paper preprint service operated by Cornell University.



\bibliographystyle{mn_new}

\label{lastpage}

\end{document}